\documentclass[%
 reprint,
groupedaddress,
 amsmath,amssymb,
 aps,
prx,
]{revtex4-2}
\usepackage[english]{babel}
\usepackage[utf8]{inputenc} \usepackage[T1]{fontenc}
\usepackage{graphicx}
\usepackage{dcolumn}
\usepackage{bm}
\usepackage{hyperref}

\usepackage{lipsum}
\usepackage{physics}
\usepackage{tikz}
\usetikzlibrary{decorations.pathmorphing}
\usepackage{bbm}
\usepackage{mathrsfs}
\usepackage{siunitx}

\makeatletter
\@ifundefined{textcolor}{}
{%
 \definecolor{BLACK}{gray}{0}
 \definecolor{WHITE}{gray}{1}
 \definecolor{RED}{rgb}{1,0,0}
 \definecolor{GREEN}{rgb}{0,1,0}
 \definecolor{BLUE}{rgb}{0,0,1}
 \definecolor{CYAN}{cmyk}{1,0,0,0}
 \definecolor{MAGENTA}{cmyk}{0,1,0,0}
 \definecolor{YELLOW}{cmyk}{0,0,1,0}
}

\begin{document}

\title{Simulating adiabatic quantum computation with a variational approach}

\author{Giuseppe Carleo}

\affiliation{Institute of Physics, École Polytechnique Fédérale de Lausanne (EPFL), CH-1015 Lausanne, Switzerland}

\author{Bela Bauer}
\author{Matthias Troyer}

\affiliation{Microsoft Azure Quantum}

\begin{abstract}
The theoretical analysis of the Adiabatic Quantum Computation protocol
presents several challenges resulting from the difficulty of simulating,
with classical resources, the unitary dynamics of a large quantum
device. We present here a variational approach to substantially alleviate
this problem in many situations of interest. Our approach is based
on the time-dependent Variational Monte Carlo method, in conjunction
with a correlated and time-dependent Jastrow ansatz. We demonstrate
that accurate results can be obtained in a variety of problems, ranging
from the description of defect generation through a dynamical phase
transition in 1D to the complex dynamics of frustrated spin-glass
problems both on fully-connected and Chimera graphs. 
\end{abstract}
\maketitle

\section{Introduction}

Adiabatic quantum computation (AQC) is a physics-inspired paradigm
to realize computations by means of controlled quantum resources.
It can be seen as an alternative to the more traditional circuit model~\cite{nielsen2010quantum},
and builds on the idea of reformulating the solution to a given
computational problem as the task of finding the ground state of an
auxiliary, problem-dependent spin Hamiltonian. The goal of the
quantum computation is then to find the requested ground-state solution
by means of an adiabatic, unitary time evolution which aims at staying
in the instantaneous ground-state of a conveniently designed driving
Hamiltonian. The resulting computing paradigm is universal in the
sense that any quantum algorithm can be reformulated in the AQC form~\cite{aharonov2007adiabatic,mizel_simple_2007}.
Moreover, a subset of AQC where the final target Hamiltonian is classical, commonly referred to as adiabatic quantum optimization (AQO), has enjoyed significant
experimental progress in recent years~\cite{johnson_quantum_2011,boixo_experimental_2013,boixo2014evidence,lanting2014entanglement},
and several applications have already been put forward~\cite{douglass_constructing_2015,venturelli2015quantum,boixo2016computational}.
The theoretical
analysis of AQC algorithms, however, is a particularly challenging
task. The largest difficulties arise because of the intrinsically
non-perturbative and strongly interacting nature of the Hamiltonians
resulting from the AQC mapping, requiring the ability to study the
unitary dynamics of an intricate many-body quantum system. As expected for a model that is universal for quantum computation, simulating
these dynamics on a classical computer
is exponentially difficult in the general case. Thus, exact numerical studies
have been mostly confined to small systems that can be accessed with exact
diagonalization~\cite{farhi2001aquantum}, and theoretical efforts
have been focused on understanding related problems that are
more easily tractable, but still give insight into AQC. On the one hand,
qualitative studies of AQO based on Quantum Monte Carlo pseudo-dynamics
(simulated quantum annealing)~\cite{santoro2002theoryof,ronnow_defining_2014,boixo2014evidence,heim_quantum_2015}
have been realized. On the other hand, large-scale unitary dynamics
has been studied for integrable models that lack frustration~\cite{dziarmaga2005dynamics,caneva2007adiabatic,zanca2016quantum},
or exhibit Hamming weight symmetry~\cite{farhi2002quantum,kong2015theperformance}.

While these classically affordable studies have provided precious
insight into the computing power of AQC, several important mechanisms
are yet to be satisfactorily addressed. The general relevance of instantons
in AQC~\cite{isakov2015understanding,jiang2016scaling}, and elucidating the many-qubit dynamics of quantum hardware~\cite{boixo2014evidence}
are two examples of largely open problems. Moreover, there are a number
of questions which cannot be assessed by equilibrium (or quasi-equilibrium)
methods. These include the study of diabatic processes~\cite{muthukrishnan2016tunneling}-- understood so far only in relatively simple models-- and the performances
of AQC in finding approximate solutions to large problems, for which
the behavior of the success probability with the total annealing time
is substantially unknown. Finally, it is known that AQC is a universal
computing paradigm only if \emph{non-stoquastic} hamiltonians are
considered \cite{kitaev_classical_2002,siu_quantum_2005,kempe_complexity_2006,biamonte_realizable_2008,bravyi2017complexity}.
The latter cannot be studied by traditional simulated quantum annealing
and pose a substantial theoretical challenge at the moment \cite{nishimori_exponential_2016,hormozi_non-stoquastic_2016}.
All these points considered, the necessity of new theoretical tools
to satisfactorily model, from first principles, AQC on large problems
appears particularly urgent.

In this paper, we introduce a variational approach to study AQC in
the unitary dynamics setting, and show that it can be used to substantially
mitigate the limitations of existing methods. Our approach is based
on the time-dependent Variational Monte Carlo method \cite{carleo2012localization,carleo2014lightcone},
suitably extended to treat time-dependent Hamiltonians. We derive
equations of motion enjoying an adiabatic variational principle which
guarantees the best unitary adiabatic evolution within the variational
manifold. We then demonstrate this approach in a rather wide variety
of problems. In particular we show that a simple two-body Jastrow
ansatz accurately describes the most common physical scenarios emerging
in AQO, such as defect generation through a dynamical phase transition
and quantum driving through a frustrated spin-glass phase. Specifically,
we first show that a high degree of precision can be obtained in the
study of a disordered 1D Ising model. Then we study a fully connected
spin glass problem, characterizing the distribution of the success
probability for the quantum algorithm. Finally, we study spin-glass
problems on a Chimera graph relevant for existing quantum
devices. On those problems, we provide an accurate determination
of the statistical properties of genuine AQC on large devices.

\section{Variational Description}

In the AQC protocol, the time evolution of the quantum system is driven
by a generic time-dependent Hamiltonian: 
\begin{eqnarray}
\mathcal{H}(t) & = & \Gamma(t;T)\mathcal{H}_{\mathrm{f}}+\left[1-\Gamma(t;T)\right]\mathcal{H}_{\mathrm{p}},\label{eq:Haqc}
\end{eqnarray}
which interpolates, for a total time $T$, between $\mathcal{H}_{\mathrm{f}}$,
whose ground state should exhibit strong quantum fluctuations in the computational (classical) basis yet be easily prepared,
and the problem Hamiltonian $\mathcal{H}_{\mathrm{p}}$.
$\Gamma(t;T)$ is a smooth function such that $\Gamma(t=0;T)=1$
and $\Gamma(t=T;T)=0$. The idea is to prepare the system in
the ground state of $\mathcal{H}(0)\equiv\mathcal{H}_{\mathrm{f}}$,
and then let the system evolve under the action of the unitary dynamics
induced by $\mathcal{H}(t)$. If the evolution is sufficiently slow,
then at time $t=T$ the ground state of the
problem hamiltonian $\mathcal{H}(T)\equiv\mathcal{H}_{\mathrm{p}}$ is prepared~\cite{kadowaki1998quantum,farhi2001aquantum,bapst2013thequantum}.

As a variational description of this dynamics, we consider a quantum
state $|\Psi(\alpha(t))\rangle$, depending on some complex-valued
and time-dependent variational parameters $\alpha(t)$. The optimal
time evolution of the variational parameters can be obtained using the Dirac-Frenkel time-dependent variational principle~\cite{dirac1930noteon,frenkel1934wavemechanics}.
In particular, the variational dynamics lies in an effective low-dimensional
Hilbert space spanned by the vectors 
\begin{equation}
\langle\boldsymbol{\sigma^{z}}|O_{k}\rangle=\frac{\langle\boldsymbol{\sigma^{z}}|\partial_{\alpha_{k}}\Psi(\alpha(t))\rangle}{\langle\boldsymbol{\sigma^{z}}|\Psi(\alpha(t))\rangle},
\end{equation}
where $\boldsymbol{\sigma^{z}}\equiv\sigma_{1}^{z},\sigma_{2}^{z},\dots\sigma_{L}^{z}$
is the chosen many-body basis. The variational generator of the dynamics
reads 
\begin{gather}
F_{\text{var}}(\boldsymbol{\sigma^{z}},t)=\frac{d}{dt}\langle\boldsymbol{\sigma^{z}}|\Psi(\alpha(t))\rangle\\
=\sum_{k}\dot{\alpha}_{k}(t)\langle\boldsymbol{\sigma^{z}}|O_{k}\rangle\langle\boldsymbol{\sigma^{z}}|\Psi(\alpha(t))\rangle,
\end{gather}
whereas the exact (Schroedinger) generator is 
\begin{equation}
F_{\text{ex}}(\boldsymbol{\sigma^{z}},t)=-i\left\langle \boldsymbol{\sigma^{z}}\right|\mathcal{H}(t)\left|\Psi(t)\right\rangle .
\end{equation}
The optimal time-evolution can be then obtained upon minimization
of the Hilbert-space distance between the variational and the exact
generator: $d^{2}(\dot{\alpha}(t))=\left\Vert F_{\text{var}}(\boldsymbol{\sigma^{z}},t)-F_{\text{ex}}(\boldsymbol{\sigma^{z}},t)\right\Vert $.
Explicit minimization leads to the following algebraic differential
equations for the variational parameters 
\begin{eqnarray}
\sum_{k^{\prime}} & \left\langle O_{k}^{\star}O_{k^{\prime}}\right\rangle _{t}^{\text{c}}\dot{\alpha}_{k^{\prime}}(t) & =-i\langle\mathcal{H}(t)O_{k}^{\star}\rangle_{t}^{c},\label{eq:t-vmc}
\end{eqnarray}
where $\left\langle AB\right\rangle _{t}^{\text{c}}=\langle AB\rangle_{t}-\langle A\rangle_{t}\langle B\rangle_{t}$
indicate connected averages, and $\langle\dots\rangle_{t}=\frac{\sum_{\sigma^{z}}\left|\Psi(\sigma^{z},t)\right|^{2}\dots}{\sum_{\sigma^{z}}\left|\Psi(\sigma^{z},t)\right|^{2}}$
are expectation values over the variational state. This approach,
in conjunction with an ansatz for correlated wave functions, leads to the time-dependent
Variational Monte Carlo (t-VMC) scheme~\cite{carleo2012localization,carleo2014lightcone}, 
which we adopt here to suitably treat time-dependent Hamiltonians.

\subsection{Variational Adiabatic Principle}

A remarkable property of the t-VMC equations, for a time-dependent
interpolating Hamiltonian (\ref{eq:Haqc}), is that they satisfy a
type of variational adiabatic principle (VAP). To clarify
this point, we introduce the rescaled time $s=t/T$, and observe that
the variational equations of motion can be written as 
\begin{eqnarray}
\frac{i}{T}\sum_{k^{\prime}}\left\langle O_{k}O_{k^{\prime}}\right\rangle _{s}^{\text{c}}\dot{\alpha}_{k^{\prime}}(s) & = & \left\langle O_{k}\mathcal{H}(s)\right\rangle _{s}^{\text{c}}.\label{eq:tdvar-rescaled}
\end{eqnarray}
In the infinite annealing time limit, the l.h.s. of this equation
vanishes provided that: the variational operators $O_{k}$ are bounded
(which is the case for the variational states we consider in the following)
and that the variational trajectories $\alpha_{k}(t)$ are sufficiently
smooth, i.e. its time derivatives stay finite during the time evolution.
Although the variational state at any time during the evolution cannot
be expected to be an eigenstate of the instantaneous Hamiltonian,
these two conditions are sufficient to guarantee that it is at least
a stationary state within the variational manifold, which represents
the closest analogue to an eigenstate available within the constraints
of the variational approximation. To see this, observe that the total
energy $E(s)=\langle\mathcal{H}(s)\rangle_{s}$ is stationary if $\partial_{\alpha_{k}}E(s)=\left\langle O_{k}\mathcal{H}(s)\right\rangle _{s}^{\text{c}}=0,$
which corresponds to the r.h.s of Eq. (\ref{eq:tdvar-rescaled}),
vanishing in the adiabatic limit. This VAP holds for any variational
state, provided it is evolved according to the t-VMC equations of
motion.

\begin{figure*}
\includegraphics[width=2\columnwidth]{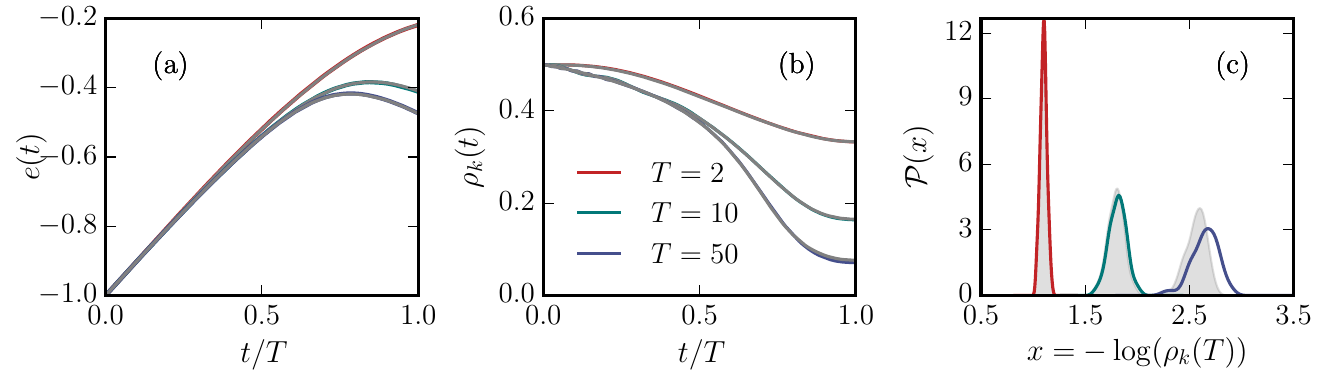}

\caption{\textbf{Variational description of quantum annealing in the disordered
1D Ising model.} Disordered-averaged energy (a) and kinks (b) densities
as a function of time $t$ and for several values of the total annealing
time $T$. T-VMC results (in color) are compared to exact ones (gray
lines, barely distinguishable). (c) Estimated probability distribution
for the logarithm of the kink-density at the end of the annealing,
t-VMC results are in color, whereas the exact distribution corresponds
to the shaded grey region. All results obtained for a system of $N=64$
qubits and represent averages over $200$ different random couplings.
\label{fig:Ising} }
\end{figure*}

\section{Applications to Adiabatic Optimization}

We now specialize our discussion to one of the most popular applications
of AQC, namely adiabatic quantum optimization. In particular we focus
on the widely used driving Hamiltonian for computational problems
in the Ising form: 
\begin{eqnarray}
\mathcal{H}(t) & = & -\Gamma(t)\sum_{i}\sigma_{i}^{x}+\left[1-\Gamma(t)\right]\sum_{i<j}V_{ij}\sigma_{i}^{z}\sigma_{i}^{z}\label{eq:H(t)}
\end{eqnarray}
where $\sigma_{i}^{\alpha}$ are Pauli matrices, the sums are defined
over a certain graph with $N$ nodes, $V_{ij}$ are bond couplings,
and $\Gamma(t)$ is a time-varying transverse field, such that $\Gamma(0)=1$
and $\Gamma(T)=0$.

In order to proceed further, an explicit ansatz for the variational
state should also be specified. 
Since the time-dependent Hamiltonian (\ref{eq:H(t)}) comprises only
few-body interactions, an amenable class of states for the variational
adiabatic evolution is the Jastrow-Feenberg expansion (JFE) \cite{feenberg_theory_1969,cevolani_protected_2015}
\begin{eqnarray}
\Psi(\sigma^{z},t) & = & \exp\left[\sum_{i}J_{i}^{(1)}\sigma_{i}^{z}+\sum_{i<j}J_{ij}^{(2)}(t)\sigma_{i}^{z}\sigma_{j}^{z}+\nonumber\right.\\
 &  & \left.+\sum_{i<j<k}J_{ijk}^{(3)}(t)\sigma_{i}^{z}\sigma_{j}^{z}\sigma_{k}^{z}+\dots\right],\label{eq:JasFee}
\end{eqnarray}
where $J^{(b)}(t)$ are complex-valued variational parameters entangling
$b$ spins and the expansions is truncated at some order $b_{\textrm{cut}}$.
This ansatz is exact for the ground state of the Hamiltonian (\ref{eq:H(t)})
in both limits $t=0$ (where $J^{(b)}(0)=0$) and $t=T$ (where the
only nonvanishing term is $J_{ij}^{(2)}(T)\propto V_{ij}$). In the
following, we will consider the two-body truncated version
of this expansion, that is, when we set $J^{(3)}=J^{(4)}\dots=0$, while
keeping one- and two-body terms. As we will demonstrate, for several
benchmark cases, this level of truncation is largely sufficient to
describe with high accuracy both qualitative and quantitative features
of AQC.

\subsection{Disordered 1D Ising Model}

\begin{figure*}
\includegraphics[width=2\columnwidth]{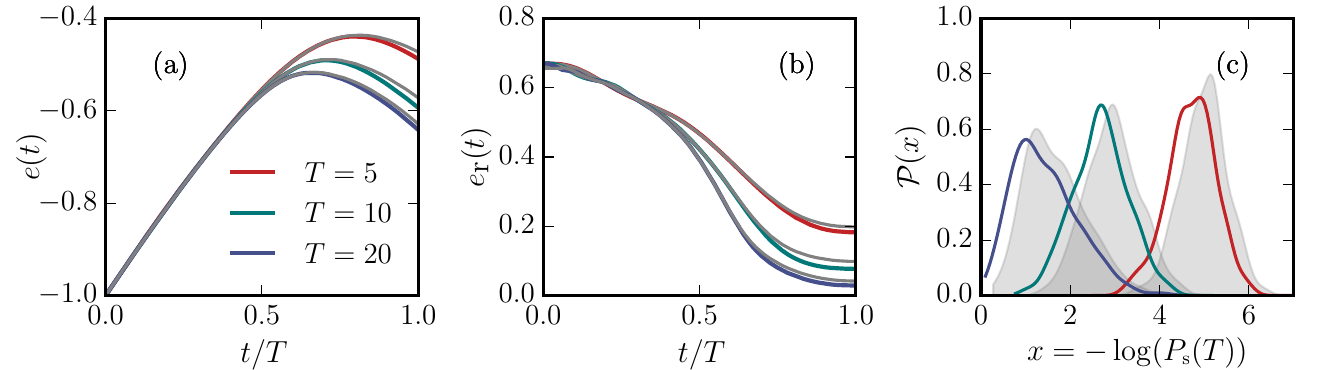} \caption{\textbf{Variational description of quantum annealing in the Sherrington-Kirkpatrick
model.} Disordered-averaged instantaneous energy (a) and residual energy (b) densities
as a function of time $t$ and for several values of the total annealing
time $T$. T-VMC results (in color) are compared with the exact ones (gray
lines). (c) Estimated success probability
at the end of the annealing. The t-VMC results are in color, whereas the
exact distribution corresponds to the shaded gray region. All results
were obtained for a system of $N=24$ qubits. \label{fig:SK} }
\end{figure*}

As a first benchmark for our approach, we study a random-bond Ising
(RI) model on a 1D lattice. This model is defined by couplings
$V_{ij}=-\delta_{i,i+1}v_{i}$, with uniformly distributed random
bonds $v_{i}\in[0,1)$. Because we restrict ourselves to nearest-neighbor couplings
in one dimension, this model is not frustrated (and remains unfrustrated
even when the couplings are chosen with random signs), and the ground
state corresponds to a ferromagnetic configuration with all the spins
aligned. Despite the simplicity of the classical ground state, the unitary time evolution induced by Eqn.~\eqref{eq:H(t)} is known to exhibit a dynamical
phase transition through a gapless point in the translationally-invariant case \cite{dziarmaga2005dynamics}.
Furthermore, the disordered model for finite $\Gamma$ is known to exhibit infinite-randomness phase transitions, which can have exponentially small gaps~\cite{fisher1999phase}.
This model is therefore a good benchmark
for the ability of our scheme to describe nonadiabatic crossings.
Another attractive feature of this model is that it can be solved exactly by means of a Jordan-Wigner transformation~\cite{dziarmaga2005dynamics,caneva2007adiabatic}.
Due to the presence of exponentially small gaps,
the time evolution, for finite values of the annealing time $T$,
will be typically dominated by diabatic effects. In particular, the
density of kinks $\rho_{k}(t)=\sum_{i}\langle1-\sigma_{i}^{z}\sigma_{i+1}^{z}\rangle_{t}/(2N)$,
which would be vanishing in the exact ground state of the final hamiltonian,
instead stays finite during the time evolution and only slowly decreases
when increasing $T$. Moreover, the probability distribution of the
kink density at $t=T$ follows a log-normal distribution, which can
be understood as a consequence of the gap distribution around the
critical point. In Fig.~\ref{fig:Ising} we show the behavior of
several quantities of interest for this model, as obtained from our
t-VMC approach (color curves) and from the exact solution (gray
curves). In Fig.~\ref{fig:Ising}-(a), we show the disordered averaged
instantaneous energy density $e(t)=\langle\mathcal{H}(t)/N\rangle_{t}$,
in Fig.~\ref{fig:Ising}-(b) the disordered-averaged instantaneous
density of kinks, and in Fig.~\ref{fig:Ising}-(c) the probability
distribution of the logarithm of the density of kinks at the end of the
annealing $(t=T)$. The latter is estimated using a kernel density method
for an ensemble of $200$ disorder realizations. Overall, the t-VMC
approach, in conjunction with the two-body JFE, reaches an impressive
level of accuracy. For the shown time-dependent quantities, the discrepancies
from the exact results are, indeed, barely visible. Only for the largest
annealing time are small deviations in the kink distribution observed,
where the variational approach tends to slightly underestimate the
kink density. The high quality of the variational results for this
problem suggests the suitability of our approach to describe the complex
many-body dynamics of a system going through a dynamical phase transition.

\subsection{Sherrington-Kirkpatrick model}

In most optimization problems of interest, couplings $V_{ij}$
can exhibit a high degree of frustration, leading to many solutions
of similar energies. To benchmark our approach also in this
case, we study spin-glass problems described by the Sherrington-Kirkpatrick
(SK) model. The SK model is defined on a fully connected graph, with
couplings $V_{ij}=\frac{v_{ij}}{\sqrt{N}}$, where $v_{ij}$ are normal
random variates. Unlike the 1d random-bond Ising model, an
exact solution for the quantum annealing of the SK model is unknown.
In the following, we therefore restrict our benchmarks to relatively
small systems, which are still manageable by exact diagonalization.
In Fig.~\ref{fig:SK}, we show several quantities of interest,
obtained from our t-VMC approach (color curves) and from the exact
solution (gray curves). In Fig.~\ref{fig:Ising}-(a), we show the
instantaneous energy density $e(t)=\langle\mathcal{H}(t)/N\rangle_{t}$
for a few values of the total annealing time $T$. In Fig.~\ref{fig:Ising}-(b),
we show the instantaneous residual energy density $e_{\mathrm{r}}(t)=\langle\mathcal{H}_{\mathrm{p}}/N\rangle_{t}-e_{0}$,
where $e_{0}$ is the ground-state energy of the problem Hamiltonian.
We also compute the success probability $P_{\textrm{s}}(T)=\langle\delta(\sum_{i<j}V_{ij}\sigma_{i}^{z}\sigma_{i}^{z}-E_{\textrm{min}})\rangle_{T}$
at the end of the annealing, and in Fig.~\ref{fig:Ising}-(c) we
show the probability distribution (averaged over 500 disorder realizations)
of $\log(P_{\textrm{s}}(T))$ both estimated on the t-VMC results (in
color) and on exact results (shaded areas). In contrast to the RI
model, we observe some deviations of the variational results from
the exact ones. In particular, variational estimates tend to overestimate
success probabilities, also resulting in lower average instantaneous
energies at finite $T$. Despite these discrepancies, however, the
overall agreement remains very satisfactory and, for example, the variational
approach provides a distribution for the success probabilities which
is very close to the exact result.

\begin{figure}[b!]
\includegraphics[width=1\columnwidth]{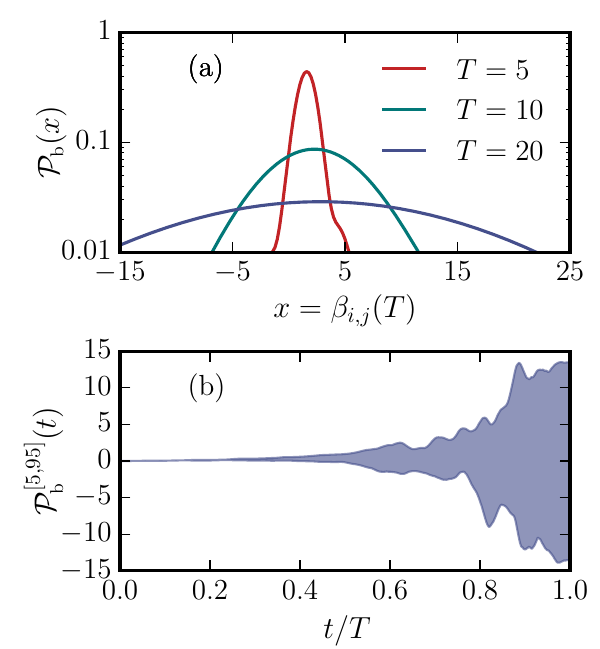} \caption{\textbf{Effective classical description of quantum annealing in the
Sherrington-Kirkpatrick model.} (a) Estimated probability distribution,
over the bonds $ij$, for effective inverse temperature at some fixed
disorder realizaton and for several values of $T$. (b) Spread of
the effective inverse temperature as a function of time,$t$, at fixed
total annealing time $T=20$. The shaded region is comprised between
the 5th and the 95th percentile of the distribution. \label{fig:SK_teff} }
\end{figure}

\subsubsection{Effective temperatures}
An intriguing point to analyze in this model is that the
two-body-truncated JFE ansatz provides in this case a natural connection
with classical annealing. This connection is best understood in terms
of the probability density 
\begin{align}
\left|\Psi(\sigma^{z},t)\right|^{2} & =\exp\left(2\sum_{i<j}\textrm{Re}\left[J_{ij}^{(2)}(t)\right]\sigma_{i}^{z}\sigma_{i}^{z}\right)\label{eq:boltzeff}
\end{align}
which directly maps onto the Boltzmann probability density of a classical
system of energy $E_{\text{eff}}(t)=-2\left(\sum_{i<j}\textrm{Re}\left[J_{ij}^{(2)}(t)\right]\sigma_{i}^{z}\sigma_{i}^{z}\right)$.
We can then conveniently introduce the time-dependent effective inverse
temperatures $\beta_{ij}^{\textrm{eff}}(t)=-2J_{ij}^{\textrm{(2)}}(t)/V_{ij}$.
The conventional approach of classical annealing with time-independent
couplings amounts to taking the temperature identical on each bond,
i.e. $\beta_{ij}^{\textrm{eff}}(t)=\beta(t)$ where $\beta(t)$ is
an increasing function of time. In the case of quantum annealing,
and using this particular ansatz, the above classical description
remains possible, but now the effective temperature can exhibit a nontrivial
bond-dependent behavior with time. In Fig.~\ref{fig:SK_teff}-(a),
we examine this behavior in the distribution (over the bonds $ij$)
of these inverse temperatures at the end of the annealing. These are
shown for a single typical instance of the SK model and for some
values of the total annealing time $T$. For small values of $T$,
the effective temperatures follow a very narrow distribution, thus closely imitating the behavior of a simple classical annealing schedule.
As the annealing time is increased, the effective temperatures exhibit
a much wider distribution. The very large effective temperatures (in absolute values) encountered for a slow quantum anneal mean that certain
bonds are effectively frozen during the annealing procedure. To further
elucidate the quantum nature of the annealing, in Fig.~\ref{fig:SK_teff}-(b)
we also show the spread of the effective inverse temperatures as a
function of time $t$ for a large total annealing time $T$. In this
case, it is possible to appreciate a strong departure from classical
annealing, signaled by the lack of a single effective temperature,
after a certain \emph{threshold} time ($t/T~0.5$), as well as the appearance of
\emph{negative} effective temperatures.

\begin{figure}
\includegraphics[width=1\columnwidth]{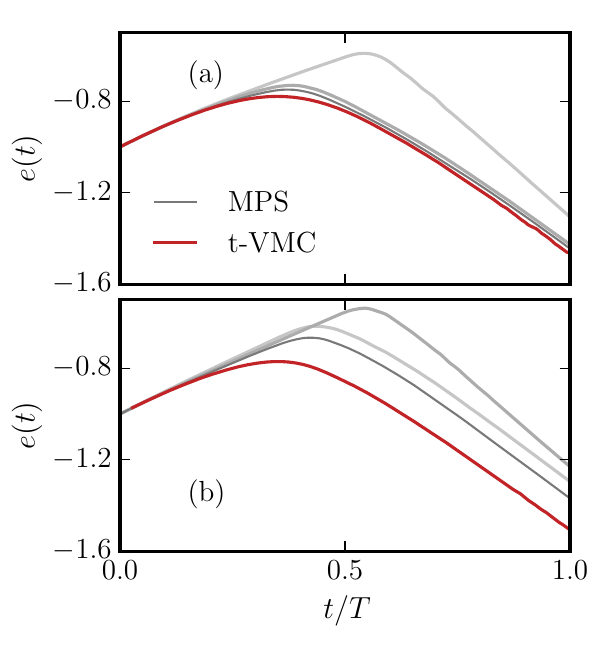}

\caption{\textbf{Comparison with Matrix Product States ansatz for Chimera graphs.}
Instantaneous energy density as a function of time $t$ and for a
fixed value of the total annealing time $T$. T-VMC results (in color)
are compared to MPS ones (gray lines, with darker colors corresponding
to more accurate results obtained with larger bond dimensions $D\in \{8,16,32\}$).
The number of qubits is $N=32$ in (a) and $N=108$
in (b), total annealing time is $T=24$ in both cases. \label{fig:mps}}
\end{figure}

\subsection{Chimera Graphs}

As a further case of study for our approach, we now concentrate on
spin-glass problems defined on Chimera graphs (CG). The latter are
two-dimensional lattices that can be experimentally realized on the
D-Wave quantum annealer. The problems defined on CG have
been the subject of several studies, e.g., \cite{okada_improving_2019,lobe_embedding_2021}.
However, the majority of previous studies have mainly focused on
the pseudodynamics of these problems, and little is known at this
stage about the physical unitary dynamics for quantum annealings on
large CG. To also benchmark our approach in this case, we
compare our results to time-dependent matrix-product state (MPS) results~\cite{fannes1992,white1992,ostlund1995,schollwock2011}.
The MPS ansatz can be interpreted as a low-entanglement approximation
to one-dimensional quantum states and yields an efficient description
of the ground states of gapped Hamiltonians in one dimension~\cite{hastings2006,hastings2007}
that exhibit an area law for the entanglement entropy~\cite{Eisert2010}.
The approximation can be systematically improved by tuning a parameter
referred to as bond dimension, $D$, which tunes the amount of entanglement that
the state can capture.

Although MPS can in principle be applied to higher-dimensional~\cite{stoudenmire2012}
as well as dynamical problems~\cite{vidal2003,daley2004,feiguin2004},
the success here becomes limited by the growth of entanglement. In
two-dimensional systems, the area law generally leads to exponential
scaling of the cost of the MPS description with the linear size of
the system. In a dynamical setup, the entanglement can grow up to
linearly in time and thus again yield exponential scaling of the cost
with time. However, in both cases, numerical experience shows that,
in many relevant cases, MPS-based simulations using highly optimized
codes can outperform exact diagonalization. The setting of adiabatic
quantum optimization seems amenable to MPS simulations because both
the initial and final states are classical and thus have a simple
MPS representation, and furthermore adiabaticity may in some cases
limit the growth of entanglement. However, additional complexity arises
if the couplings are non-local; for a discussion of this issue, see~\cite{bauer2015}, which also discusses the application of higher-dimensional
generalizations of MPS, such as PEPS~\cite{hieida1999numerical,nishino2001,gendiar2003,verstraete2004},
to adiabatic quantum optimization. We note that the variational class
of states used in our t-VMC simulations is closely related to complete-graph
tensor network states~\cite{marti2010,changlani2009}.

\begin{figure}[t!]
\includegraphics[width=1\columnwidth]{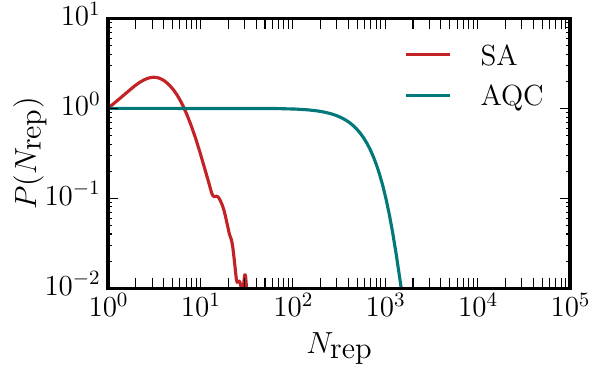}

\caption{\textbf{Probability distribution for the number of repetitions on
Chimera graphs.} (Red curve) Classical Simulated Annealing for a number
of sweeps $N_{s}=1000$. (Right) Adiabatic Quantum Computing at the
annealing time $T=20$. In both cases the probability density is normalized
by the value at $N_{\textrm{rep}}=1$. The Chimera graph considered
has $72$ qubits, and $500$ disorder realizations have been used
to produce the estimates for the probability density. \label{fig:chimeradist}
}
\end{figure}

We perform numerical simulations based on MPS for the smallest non-trivial
Chimera graph, which consists of $N=32$ sites, as well as a larger
graph of $N=108$ sites. In Fig.~\ref{fig:mps} we show a comparison
between the instantaneous energy density obtained by t-VMC
(red lines) and by the MPS (where darker curves correspond to more
accurate results obtained with increasing bond dimensions). We find
that the t-VMC approach yields lower instantaneous and final energies
in both cases, compared to the case of MPS with the largest bond dimension we considered here, up to $D=32$.
For a system of 32 qubits, the MPS obtains a good approximation
for a bond dimension of $D=32$. On the largest system with 108 qubits 
they perform drastically worse when keeping the bond dimension constant, at the same value adopted on the smaller system. 
This is because mapping the non-planar two-dimensional chimera graph onto a chain leads to
a very non-local entanglement structure in this chain, thus requiring
an exponentially large bond dimension for an accurate description.
We have not attempted scaling the bond dimension beyond the value reported in the 
text, and we leave a more systematic analysis to future studies. 

\subsubsection{Efficiency of the adiabatic protocol}
A convenient way to characterize the efficiency of the AQC protocol
to find the ground state of the spin glass problem is to introduce
the average number of repetitions required to obtain the solution
with a fixed reference probability $p_{0}$. In particular, we consider
$N_{\textrm{rep}}=\log(1-p_{0})/\log(1-P_{\textrm{s}}(T))$, with
$p_{0}=0.99$. In Fig.~\ref{fig:chimeradist} we show the behavior
of the probability distribution for $N_{\textrm{rep}}$ at some fixed,
typical value of the total annealing time $T$. In the same Figure,
we also show the average number of repetition for classical Simulated
Annealing, on the same problems analyzed for the quantum case. The
two distributions are qualitatively different. On the one hand, the classical
annealing is relatively more compact, while exhibiting a behavior
compatible with power-law tails, as found in Ref.~\cite{steiger2015heavytails}.
On the other hand, the AQC distribution is substantially more flat,
exhibiting a characteristic plateau over several orders of magnitude.
While a detailed analysis of the tails is beyond the scope of this paper, we can, however, already see that the behavior of the
unitary AQC shows some differences from the SQA one. In particular,
the SQA distribution, also analyzed in Ref.~\cite{steiger2015heavytails},
does not exhibit the large plateau we observe here. This plateau might
therefore be a characteristic trait of the unitary dynamics AQC, as
opposed to the quantum Monte Carlo pseudo-dynamics studied until now.

\section{Outlook and Discussion}

We have introduced a variational approach to study the unitary time
evolution of interacting spin systems performing a quantum annealing.
Several problems previously difficult to study with existing approaches
can now be addressed thanks to our variational method. These range
from the effect of diabatic transitions to the effort of understanding
the possible origins of quantum speed-up. We have shown here how a
simple two-body Jastrow ansatz is sufficient to accurately describe
most of the properties of interest, when the AQC Hamiltonian involves
only few-body interactions. The extension of our approach to study
problem Hamiltonians with several-body interactions is feasible, provided that
a sensible wave-function ansatz is chosen. A particularly interesting
class of variational states is the more recently introduced neural network quantum states~\cite{carleo2016solving}, which could be used to
study AQC in the most challenging applications. 

\paragraph*{Note Added} 
This work was completed in 2016 and presented at the APS March Meeting in its final form~\cite{carleo_accurate_2016}. A preprint with minimal changes was published in 2024.   
\begin{acknowledgments}
We acknowledge the discussions with A. Scardicchio, G. Semerjian, D. Venturelli,
and F. Zamponi. This work was supported by the European Research Council
through ERC Advanced Grant SIMCOFE, by the Swiss National Science
Foundation through NCCR QSIT, by Microsoft Research, and by a grant
from the Swiss National Supercomputing Centre (CSCS) under project
ID s686. This article is based on work supported in part by ODNI,
IARPA via MIT Lincoln Laboratory Air Force Contract No. FA8721-05-C-0002.
The views and conclusions contained herein are those of the authors
and should not be interpreted as necessarily representing the official
policies or endorsements, either expressed or implied, of ODNI, IARPA,
or the U.S. Government. The U.S. Government is authorized to reproduce
and distribute reprints for Governmental purpose not-withstanding
any copyright annotation thereon. 
\end{acknowledgments}

\bibliography{biblio}

\end{document}